\documentclass[aps,prl,twocolumn,showpacs,floatfix]{revtex4}

\usepackage{graphicx} 
\usepackage{bm}

\begin{document}

\title{Theory of the high-frequency chiral optical response in a
\boldmath $p_x+ip_y$ superconductor}

\author{Victor M.~Yakovenko} 

\affiliation{Department of Physics, University of
  Maryland, College Park, Maryland 20742-4111, USA}

\date{\bf cond-mat/0608148, v.4 February 25, 2007}


\begin{abstract} 
  The optical Hall conductivity and the polar Kerr angle are
  calculated as functions of temperature for a two-dimensional chiral
  $p_x+ip_y$ superconductor, where the time-reversal symmetry is
  spontaneously broken.  The theoretical estimate for the polar Kerr
  angle agrees by the order of magnitude with the recent experimental
  measurement in $\rm Sr_2RuO_4$ by Xia {\it et al.} [PRL {\bf 97},
  167002 (2006)].  The theory predicts that the Kerr angle is
  proportional to the square of the superconducting energy gap and is
  inversely proportional to the cube of frequency, which can be
  verified experimentally.
\end{abstract} 

\maketitle

Xia {\it et al.}\ \cite{Xia06} recently reported experimental
observation of the polar Kerr effect in the superconducting state of
$\rm Sr_2RuO_4$.  In the absence of an external magnetic field,
reflected light shows rotation of polarization, which is a clear
signature of the spontaneous time-reversal-symmetry breaking in the
superconducting state \cite{PhysToday}.  Previous muon spin relaxation
measurements \cite{Luke} suggested the time-reversal symmetry is
broken in $\rm Sr_2RuO_4$, but the polar Kerr experiment \cite{Xia06}
gives a much more convincing evidence for this remarkable effect.

$\rm Sr_2RuO_4$ consists of weakly coupled two-dimensional (2D)
metallic layers.  It was proposed theoretically that the
superconducting pairing in this material is spin-triplet
\cite{Baskaran96} and has the chiral $p_x+ip_y$ symmetry
\cite{Rice95}.  Such an order parameter breaks the time-reversal
symmetry and is analogous to the 2D superfluid $^3$He-A
\cite{Volovik88}.  There is substantial experimental evidence in favor
of the spin triplet and odd orbital symmetry of the superconducting
pairing in $\rm Sr_2RuO_4$ \cite{Mackenzie03}, which includes
measurements of the spin susceptibility \cite{Ishida98} and the
Josephson effect \cite{Nelson04} (see, however, an alternative
interpretation \cite{Mazin05}).  On the other hand, the chiral
character was not so well established experimentally (see Ref.\
\cite{Sengupta02} for interpretation of tunneling measurements).

Although the experimental demonstration \cite{Xia06} of the
spontaneous polar Kerr effect is very convincing, a theory of this
effect for chiral superconductors is not well developed.  Theories
\cite{Joynt91,Yip92} concluded that there is no chiral term in the
single-particle response of a $p_x+ip_y$ superconductor, although
Fig.\ 1 of Ref.\ \cite{Joynt91} shows a non-zero Kerr effect for a
different state, and Ref.\ \cite{Yip92} found some chiral response
from collective excitation.  On the other hand, Ref.\ \cite{Volovik88}
obtained the intrinsic quantum Hall effect in the single-particle
response of a $p_x+ip_y$ superconductor, which was then studied in
much detail in Ref.\ \cite{Furusaki01}.  Following Refs.\
\cite{Xia06,Joynt91,White-Geballe}, the polar Kerr angle $\theta_K$
can be expressed in terms of the imaginary part of the ac Hall
conductivity $\sigma_{xy}''(\Omega)$ at a frequency $\Omega$
\begin{equation}
  \theta_K = \frac{4\pi}{n(n^2-1)\Omega d}\,\sigma_{xy}''(\Omega),
\label{Kerr}
\end{equation}
where $n$ is the refraction coefficient.  We write Eq.\ (\ref{Kerr})
in terms of the 2D Hall conductivity $\sigma_{xy}$ per one layer,
which is related to the 3D one via the interlayer distance $d$.  The
natural dimensional scale for the 2D $\sigma_{xy}$ is $e^2/h$.

In this paper, we calculate the ac Hall conductivity
$\sigma_{xy}(\Omega)$ at a finite frequency $\Omega$ as a function of
temperature $T$ for a $p_x+ip_y$ superconductor.  We generalize the
method of Refs.\ \cite{Volovik88,Furusaki01} and obtain the
Chern-Simons-like term in the effective action at finite $T$ and
$\Omega$.  In the intermediate calculations, we set the Planck
constant to unity $\hbar\to1$, but restore it in the final results.
The Lagrangian of electrons $L=i\partial_t-H$ (where $H$ is the
Hamiltonian) for the 2D $p_x+ip_y$ superconductor has the form
\cite{Volovik88}
\begin{equation}
  L=\left( 
  \begin{array}{cc}
  i\partial_t + \bm\nabla^2/2m + \mu 
  & i(\bm\nabla\cdot\bm\Psi+\bm\Psi\cdot\bm\nabla)/2 \\
  i(\bm\nabla\cdot\bm\Psi^*+\bm\Psi^*\cdot\bm\nabla)/2
   & i\partial_t - \bm\nabla^2/2m - \mu 
  \end{array}  \right).
\label{L0}
\end{equation}
Here $\partial_t$ and $\bm\nabla=(\partial_x,\partial_y)$ represent
time and space derivatives, $m$ and $\mu$ are the mass and the
chemical potential of the electrons.  We assume the parabolic
dispersion law $\varepsilon(\bm p)=\bm p^2/2m-\mu$, where $\bm
p=(p_x,p_y)$ is the electron momentum.  The superconducting order
parameter is $\bm\Psi=\Delta_x\hat x+i\Delta_y\hat y$, where $\hat x$
and $\hat y$ are the unit vectors in the $x$ and $y$ directions.
Because of the square symmetry in $\rm Sr_2RuO_4$, we have
$\Delta_x=\Delta_y$, but it is convenient to label the two components
of $\bm\Psi$ differently for the clarity of calculations.  In momentum
representation, Eq.\ (\ref{L0}) can be written as
\begin{equation}
  L = i\omega - \varepsilon(\bm p)\tau_3 - p_x\Delta_x\tau_1
  + p_y\Delta_y\tau_2,
\label{L0-tau}
\end{equation}
where the Pauli matrices $\tau$ act on the spinor $[\psi(\bm
p),\psi^+(-\bm p)]$ consisting of the particle and hole operators.  We
do not write the spin indices of electrons explicitly.  The two spin
components give the same contributions to the Hall conductivity, so
the final results should be multiplied by 2.  However, by introducing
electron and hole operators, we artificially doubled the number of
components, so the final result should be divided by 2
\cite{Volovik88}.  Thus, we can obtain the correct result by
considering just one spin component, as implied in Eq.\
(\ref{L0-tau}).  In Eq.\ (\ref{L0-tau}), we use the Matsubara
frequency $i\omega$, because we will be doing calculations at a finite
temperature.  From Eq.\ (\ref{L0-tau}), the Green function of
electrons $G=L^{-1}$ is:
\begin{equation}
  G(\vec p) = -\frac{i\omega + \varepsilon(\bm p)\tau_3 +
  p_x\Delta_x\tau_1 - p_y\Delta_y\tau_2} {\omega^2 + E^2(\bm p)},
\label{G0-tau}
\end{equation}
where $\vec p=(\omega,\bm p)$ is the three-component
frequency-momentum vector, and $E(\bm p)=\sqrt{\varepsilon^2(\bm
p)+p_x^2\Delta_x^2+p_y^2\Delta_y^2}$ is the electron dispersion in the
superconducting state.

To calculate electromagnetic response of the system, we introduce the
electromagnetic potentials $\vec A=(A_0,A_x,A_y)$ by using the long
derivatives $-i\bm\nabla\mp e{\bm A}/c$ and $-i\partial_t\pm eA_0$ in
the diagonal terms of Eq.\ (\ref{L0}), where $e$ is the electron
charge, and $c$ is the speed of light \cite{Volovik88}.  We also
assume that the superconducting order parameter has a
space-time-dependent phase $\varphi$, so that it can be written as
$\bm\Psi=e^{i\varphi}\bm\Psi_0$, where $\bm\Psi_0$ is uniform with the
real $\Delta_x$ and $\Delta_y$.  We will see that the effective action
depends only on gradients of $\varphi$.  To the first order in $\vec
A$ and $\varphi$, we find the following addition to the Lagrangian:
\begin{equation}
  \Gamma=\left(
  \begin{array}{cc}\displaystyle 
  -eA_0 - i\frac{e{\bm A}\cdot\bm\nabla}{mc}
  &\displaystyle 
  -\frac{\bm\nabla\cdot\bm\Psi_0\,\varphi
  +\varphi\,\bm\Psi_0\cdot\bm\nabla}{2} 
\\ \displaystyle
  \frac{\bm\nabla\cdot\bm\Psi_0^*\,\varphi
  +\varphi\,\bm\Psi_0^*\cdot\bm\nabla}{2}
   & \displaystyle eA_0 - i\frac{e{\bm A}\cdot\bm\nabla}{mc}
  \end{array}  \right).
\label{L1}
\end{equation}
The Fourier transform of Eq.\ (\ref{L1}) can be written as
\begin{equation}
  \Gamma = -eA_0\tau_3 + e{\bm A}\cdot\bm p/mc
   - \varphi p_x\Delta_x\tau_2 - \varphi p_y\Delta_y\tau_1.
\label{Gamma}
\end{equation}
Here the variables $\vec A$ and $\varphi$ are assumed to be functions
of the Fourier variable $\vec q=(\Omega,q_x,q_y)$.  Thus, the vertex
$\Gamma(\vec q,\vec p)$ (\ref{Gamma}) is a function of two vector
arguments.

The effective action of the system to the second order in $\Gamma$ is
\begin{equation}
  S=\frac12\sum_{\vec q,\vec p}{\rm Tr}\,
  \Gamma(\vec q,\vec p)\,G(\vec p+\vec q/2)\,
  \Gamma(-\vec q,\vec p)\,G(\vec p-\vec q/2).
\label{S}
\end{equation}
Substituting Eqs.\ (\ref{G0-tau}) and (\ref{Gamma}) into Eq.\
(\ref{S}), we write $S$ in the form
\begin{equation}
  S=\sum_{\vec q,\vec p}\,\frac{C_1}{C_2},
\label{S'}
\end{equation}
where the denominator is 
\begin{equation}
  C_2 = [(\omega+\Omega/2)^2 + E^2(\bm p+\bm q/2)]
  [(\omega-\Omega/2)^2 + E^2(\bm p-\bm q/2)],
\label{C2}
\end{equation}
and the numerator is
\begin{widetext}
\begin{eqnarray}
  C_1 =\frac12 &{\rm Tr}&  
  [-eA_0(\vec q)\tau_3+ep_xA_x(\vec q)/mc+ep_yA_y(\vec q)/mc
  -\varphi(\vec q)p_x\Delta_x\tau_2-\varphi(\vec q)p_y\Delta_y\tau_1]
\nonumber \\
  &\times& [i(\omega+\Omega/2)+\varepsilon({\bm p}+{\bm q}/2)\tau_3
  +(p_x+q_x/2)\Delta_x\tau_1 -(p_y+q_y/2)\Delta_y\tau_2]
\nonumber \\
  &\times& [-eA_0(-\vec q)\tau_3+ep_xA_x(-\vec q)/mc+ep_yA_y(-\vec q)/mc
  -\varphi(-\vec q)p_x\Delta_x\tau_2-\varphi(-\vec q)p_y\Delta_y\tau_1]
\nonumber \\
  &\times& [i(\omega-\Omega/2)+\varepsilon({\bm p}-{\bm q}/2)\tau_3
  +(p_x-q_x/2)\Delta_x\tau_1 -(p_y-q_y/2)\Delta_y\tau_2]
\label{C1}
\end{eqnarray}
\end{widetext}
The calculation of the effective action (\ref{S}) is conceptually
similar to the calculation of electromagnetic response in the BCS
theory of superconductivity \cite{Joynt91,Mahan}.  However, we focus
only on obtaining the Chern-Simons-like term responsible for
$\sigma_{xy}$ \cite{Volovik88,Furusaki01}.  Picking the $A_0$ term
from the first factor in Eq.\ (\ref{C1}) and the $A_x$ or $A_y$ term
from the third factor, we obtain a non-zero contribution after taking
trace over the $\tau$ matrices.  The same procedure works for the
$A_x$ or $A_y$ term from the first factor and the $A_0$ term from the
third factor.  Combining these terms and changing the variable of
integration $\vec q\to-\vec q$ in the latter term, we obtain one
contribution to $C_1$
\begin{equation}
  C_1^{(a)}= A_0(\vec q)\,[-q_yA_x(-\vec q)p_x^2 + q_xA_y(-\vec q)p_y^2]
  \,\frac{2i\Delta_x\Delta_ye^2}{mc}.
\label{C1a-q}
\end{equation}
In deriving (\ref{C1a-q}), we omitted the terms proportional to the
product $p_xp_y$, which would vanish after integration over $p_x$ and
$p_y$.  The integration over momentum $\bm p$ in Eq.\ (\ref{S'}) is
concentrated near the Fermi surface, so we can replace $p_x^2\to
p_F^2/2$ and $p_y^2\to p_F^2/2$ in Eq.\ (\ref{C1a-q}), because
$p_x^2+p_y^2\approx p_F^2$.  Making the Fourier transform of Eq.\
(\ref{C1a-q}) to the coordinate space, we find
\begin{equation}
  C_1^{(a)}= A_0(\partial_yA_x-\partial_xA_y)\frac{\Delta_0^2e^2}{mc},
\label{C1a}
\end{equation}
where $\Delta_0=\Delta_xp_F=\Delta_yp_F$ is the energy gap at the
Fermi level.

Picking the last two terms in the first factor in Eq.\ (\ref{C1}) and
the $A_x$ and $A_y$ terms in the third factor or vice versa, we obtain
another contribution to $C_1$:
\begin{equation}
  C_1^{(b)}= i\Omega\varphi(\vec q)
  [-q_yA_x(-\vec q)p_x^2 + q_xA_y(-\vec q)p_y^2]
  \frac{\Delta_x\Delta_ye}{mc}.
\label{C1b-q}
\end{equation}
Replacing the Matsubara frequency by the real frequency
$i\Omega\to\Omega$ in Eq.\ (\ref{C1b-q}) and Fourier-transforming to
the coordinate space, we find
\begin{equation}
  C_1^{(b)}= \partial_t\varphi(\partial_yA_x-\partial_xA_y)
  \frac{\Delta_0^2e}{2mc},
\label{C1b}
\end{equation}
Combining the contributions (\ref{C1a}) and (\ref{C1b}) to $C_1$ and
substituting into (\ref{S'}), we find a Chern-Simons-like term in the
effective action \cite{Volovik88,Furusaki01}
\begin{equation}
  S_{\rm CS}=\sigma_{xy}\int dt\,dx\,dy\, 
  (A_0+\partial_t\varphi/2e)(\partial_yA_x-\partial_xA_y)/c,
\label{CS}
\end{equation}
where
\begin{equation}
  \sigma_{xy}=\frac{\Delta_0^2e^2}{m} 
  \sum_{\vec p}\,\frac{1}{C_2}
\label{s_xy}
\end{equation}
is the effective Hall conductivity.  Indeed, taking the variational
derivative of Eq.\ (\ref{CS}), we find electric current
\begin{equation}
  \bm j=c\,\frac{\delta S_{\rm CS}}{\delta\bm A}
  =\sigma_{xy}\left[\bm E-
  \frac{1}{2e}\partial_t\left(\bm\nabla\varphi-\frac{2e}{c}\bm A\right)
  \right]\times\hat z,
\label{j_xy}
\end{equation}
where $\bm E=-\bm\nabla A_0-\partial_t\bm A/c$ is the electric field,
and the last term in Eq.\ (\ref{j_xy}) is proportional to the time
derivative of the London supercurrent $\bm j_s=(\rho_s
e/2m)[\bm\nabla\varphi-(2e/c)\bm A]$.  Obtaining a self-consistent
equation of motion for the superconducting phase $\varphi$ is a
complicated problem \cite{UFN}.  However, one may argue that the
supercurrent contribution in Eq.\ (\ref{j_xy}) is ineffective at high
frequencies, so the last term can be omitted, and we obtain the
standard relation for the Hall conductivity $\bm j=\sigma_{xy}\bm
E\times\hat z$.  The Chern-Simons-like term (\ref{CS}) was derived in
Ref.\ \cite{Volovik88} at $T=0$ and in Ref.\ \cite{Furusaki01} near
$T_c$ via the Ginzburg-Landau expansion.  Notice that is does not have
the component $A_x\partial_tA_y$, which is present in the standard
Chern-Simons term $\epsilon^{\mu\nu\lambda}A_\mu F_{\nu\lambda}$.
Nevertheless, Eq.\ (\ref{CS}) is gauge-invariant, because the gauge
transformation of $A_0$ in the first factor is compensated by
transformation of the superconducting phase $\varphi$, and the last
factor is manifestly gauge-invariant \cite{Volovik88}.

Now we substitute Eq.\ (\ref{C2}) into Eq.\ (\ref{s_xy}) and derive an
explicit expression for the Hall conductivity.  To obtain optical
response at a finite temperature, we do analytical continuation from
the Matsubara to real frequencies, which is well known in the BCS
theory \cite{Mahan}.  We take the limit $\bm q\to0$ while keeping
finite frequency, as appropriate for optical response, and find
\begin{eqnarray}
  \sigma_{xy}(\Omega)&=&\frac{e^2\Delta_0^2}{8\pi}
  \int_{-\infty}^\infty d\varepsilon\, \frac{1-2n(E/T)}{E^2}
\label{s_xy-w} \\
  &&\times\left(
  -\frac{1}{\Omega +i\gamma -2E}
  +\frac{1}{\Omega+i\gamma+2E}\right),
\nonumber
\end{eqnarray}
where $E=\sqrt{\varepsilon^2+\Delta_0^2}$, $n(E/T)$ is the Fermi
distribution function, $\Omega$ is the real frequency, $\gamma$ is a
relaxation rate, and we replaced the integration over
$dp_x\,dp_y/(2\pi)^2m$ by the integration over $d\varepsilon/2\pi$.
First we take the dc limit $\Omega\to0$ at zero temperature $T\to0$ in
Eq.\ (\ref{s_xy-w}) and find
\begin{equation}
  \sigma_{xy}^{\rm dc}=\frac{e^2}{4\pi}=\frac{e^2}{2h},
\label{s_xy^0}
\end{equation}
where we restored the dimensional factor $\hbar=h/2\pi$ in the
denominator.  Eq.\ (\ref{s_xy^0}) demonstrates the half-integer
quantum Hall conductivity in agreement with Ref.\ \cite{Volovik88}.
As discussed in Ref.\ \cite{Furusaki01}, it is difficult to measure
the dc Hall conductivity experimentally because of screening by
supercurrents.

At a finite temperature $T$, the dc Hall conductivity
$\sigma_{xy}^{\rm dc}(T)=(e^2/2h)f_d(T)$ is reduced by the factor
\begin{equation}
  f_d(T)=\frac{\Delta_0^2}{2} \int_{-\infty}^\infty
  d\varepsilon\,\frac{1-2n(E/T)}{E^{3/2}}.
\label{s_xy-T}
\end{equation}
The factor $f_d(T)$ interpolates between 1 at $T=0$ and 0 at $T=T_c$
and behaves as $f_d(T)\propto\Delta\propto\sqrt{T_c-T}$ near $T_c$.
The same factor (\ref{s_xy-T}) describes temperature dependence of the
quantum Hall effect in the magnetic-field-induced spin-density-wave
(FISDW) state of the quasi-one-dimensional organic conductors
$\rm(TMTSF)_2X$ \cite{FISDW,Goan98}.  Eq.\ (\ref{s_xy-T}) represents
the dynamic limit of the dc electromagnetic response
\cite{FISDW,Goan98}.  If the limit $\Omega\to0$ is taken in Eq.\
(\ref{s_xy}) first and then $\bm q\to0$, that would generate the
static limit $f_s(T)$ for the reduction function, which has the same
temperature dependence as the London superfluid density $\rho_s(T)$,
particularly $f_s(T)\propto\Delta^2\propto(T_c-T)$ near $T_c$ (see
discussion in Sec.\ VI of Ref.\ \cite{Goan98}).  Ref.\
\cite{Furusaki01} obtained the static limit for $\sigma_{xy}^{\rm dc}$
near $T_c$ by doing the Ginzburg-Landau expansion.

Now we calculate the imaginary part of the Hall conductivity
$\sigma_{xy}''(\Omega)$ at a high frequency $\Omega\gg\Delta_0$.  One
contribution originates from the pole at $\Omega=2E$ in Eq.\
(\ref{s_xy-w}).  This term represents creation of an electron pair
above the energy gap $\Delta_0$ or a hole pair below the gap by
absorption of a photon with the frequency $\Omega$.  By integrating
over $\varepsilon\approx\pm E$ in the vicinity of the resonance, we
find
\begin{equation}
  \sigma_{xy}''(\Omega)=\frac{e^2\Delta_0^2}{2\Omega^2}
  =\frac{e^2}{2\hbar}\left(\frac{\Delta_0}{\hbar\Omega}\right)^2.
\label{s_xy-1}
\end{equation}
Calculating this term, we set $E^2=(\Omega/2)^2$ in the denominator of
the first factor in Eq.\ (\ref{s_xy-w}) and put $n(E/T)=0$, because
$\Omega\gg T$.  We observe that Eq.\ (\ref{s_xy-1}) does not depend on
the relaxation rate $\gamma$ and is reduced relative to Eq.\
(\ref{s_xy^0}) by the factor $(\Delta_0/\hbar\Omega)^2/2\pi$.  The
temperature dependence of $\sigma_{xy}''(\Omega)$ is given by
$\Delta_0^2(T)$.

There is another contribution to the integral (\ref{s_xy-w})
originating from the peak in the density of states at
$E\approx\Delta_0$.  By changing the variable on integration from
$\varepsilon$ to $E$, we rewrite Eq.\ (\ref{s_xy-w}) as follows
\begin{equation}
  \sigma_{xy}(\Omega)=-\frac{e^2\Delta_0^2}{\pi}
  \int_{\Delta_0}^\infty dE\, \frac{1-2n(E/T)}{\sqrt{E^2-\Delta_0^2}}
  \frac{1}{(\Omega+i\gamma)^2-4E^2}.
\label{s_xy-2}
\end{equation}
Integral (\ref{s_xy-2}) over $dE$ is logarithmic between $\Delta_0$
and $\Omega$.  For simplicity, we consider low temperatures, where
$n(E/T)\approx0$, and find the following contribution
$\tilde\sigma_{xy}$ to the Hall conductivity
\begin{eqnarray}
  &&\tilde\sigma_{xy}(\Omega)\approx
  -\frac{e^2}{\pi}\frac{\Delta_0^2}{(\Omega+i\gamma)^2}
  \ln\left(\frac{\Omega}{\Delta_0}\right),
\nonumber \\
  &&\tilde\sigma_{xy}''(\Omega)\approx
  -\frac{4e^2}{h}\frac{\Delta_0^2\,\hbar\gamma}{(\hbar\Omega)^3}
  \ln\left(\frac{\hbar\Omega}{\Delta_0}\right).
\label{s_xy-2''}
\end{eqnarray}
Eq.\ (\ref{s_xy-2''}) is reduced relative to Eq.\ (\ref{s_xy-1}) by
the factor $\gamma/\Omega$ and is enhanced by the factor
$\ln(\hbar\Omega/\Delta_0)$.  Using the numbers from Ref.\
\cite{Xia06} and given below, we conclude that the reduction of Eq.\
(\ref{s_xy-2''}) is much greater than the enhancement, so we focus
only on Eq.\ (\ref{s_xy-1}).

Substituting Eq.\ (\ref{s_xy-1}) into Eq.\ (\ref{Kerr}), we find the
Kerr angle
\begin{equation}
  \theta_K = \frac{2\pi}{n(n^2-1)}\,\frac{e^2}{d}\,
  \frac{\Delta_0^2}{(\hbar\Omega)^3}
  =\frac{\alpha}{n(n^2-1)}\,\frac{\lambda}{d}\,
  \frac{\Delta_0^2}{(\hbar\Omega)^2},
\label{Kerr'}
\end{equation}
where $\alpha=e^2/\hbar c=1/137$ is the fine structure constant, and
$\lambda$ is the wavelength of light.  Using the interlayer distance
$d=1.3$ nm \cite{Mackenzie03} and the values $n(n^2-1)=3$ and
$\lambda=1550$ nm from Ref.\ \cite{Xia06}, we find that the first two
factors in Eq.\ (\ref{Kerr'}) give 2.9.  Using the BCS formula
$\Delta_0=1.76\,k_BT_c=0.23$ meV for $T_c=1.5$ K \cite{Xia06} and
$\hbar\Omega=hc/\lambda=0.8$ eV, we find that the last factor in Eq.\
(\ref{Kerr'}) is $(\Delta_0/\hbar\Omega)^2=8\times10^{-8}$.  The
resultant Kerr angle (\ref{Kerr'}) is $\theta_K=230$ nanorad.  This
estimate is 3.6 times greater than the experimentally observed value
of 65 nanorad \cite{Xia06}.  The experimental Kerr angle may be
reduced relative to the theoretical estimate for variety of reasons.
For example, the effective value of $\Delta_0$ at high energies may be
lower than at the Fermi level.  We conclude that the theoretical
formula (\ref{Kerr'}) reasonably agrees with the experiment by the
order of magnitude.

A different theoretical formula with $\theta_K\propto\Delta_0$ was
proposed phenomenologically in Ref.\ \cite{Xia06} motivated by the
experimental temperature dependence of $\theta_K(T)$.  On the other
hand, our formula (\ref{Kerr'}) gives $\theta_K\propto\Delta_0^2$.
The error bars in the experiment \cite{Xia06} are quite big, so
deciding between the linear or quadratic dependences of $\theta_K$ on
$\Delta_0$ may require more precise measurements.  The appearance of
$\Delta_0^2$ in Eqs.\ (\ref{s_xy-1}) and (\ref{Kerr'}) is quite
natural, originating from the product $\Delta_x\Delta_y$, which
changes sign when the chirality of the order parameter changes from
$p_x+ip_y$ to $p_x-ip_y$, as observed experimentally \cite{Xia06}.
This product can be also written as the vector
$\bm\Psi\times\bm\Psi^*$ pointing along $\hat z$ \cite{Furusaki01},
which is consistent with Eq.\ (\ref{j_xy}).  It would be very
interesting to verify experimentally the $\Omega^{-3}$ frequency
dependence of the Kerr angle predicted by Eq.\ (\ref{Kerr'}).

Experiments indicate that the superconducting gap may have the
so-called horizontal lines of nodes in $\rm Sr_2RuO_4$ (see Ref.\
\cite{Sengupta02} and references therein).  In this case, $\Delta_0$
should be considered a function of the electron momentum $p_z$
perpendicular to the layers: $\Delta_0\to\Delta_0\cos(p_zd)$ for
triplet pairing \cite{Sengupta02} or $\Delta_0\to\Delta_0\sin(p_zd)$
for singlet pairing \cite{Mazin05}.  In both cases, averaging
$\Delta_0^2(p_z)$ over $p_z$ generates an additional factor 1/2 in
Eq.\ (\ref{Kerr'}) and no changes in Eq.\ (\ref{s_xy^0}).  Thus,
although experiment \cite{Xia06} directly proves the chiral character
of the superconducting pairing in $\rm Sr_2RuO_4$, it does not
discriminate between triplet pairing and the chiral singlet pairing
$(p_x+ip_y)\sin(p_zd)$ proposed in Ref.\ \cite{Mazin05}.

In conclusion, we derived the Chern-Simons-like term in the effective
action of the 2D chiral $p_x+ip_y$ superconductor, generalizing
previous results \cite{Volovik88,Furusaki01} to finite frequency and
temperature.  The resultant dc Hall conductivity has the half-quantum
value $\sigma_{xy}=e^2/2h$ at $T=0$ \cite{Volovik88}, but is reduced
at a finite temperature by the factor (\ref{s_xy-T}).  We derived Eq.\
(\ref{s_xy-1}) for the imaginary part of the optical Hall conductivity
and Eq.\ (\ref{Kerr'}) for the polar Kerr angle, which agrees by the
order of magnitude with the recent experimental measurement in $\rm
Sr_2RuO_4$ by Xia {\it et al.}\ \cite{Xia06}.  Eq.\ (\ref{Kerr'})
predicts that the Kerr angle is proportional to the square of the
superconducting energy gap and is inversely proportional to the cube
of frequency, which can be verified experimentally.  The derivation
may be also relevant for the finite-temperature Chern-Simons theories
in high-energy physics (see references in \cite{Universe,Sengupta00}).

We thank S.~Tewari, S.~Das Sarma, and K.~Sengupta for useful
suggestions.



\end{document}